\documentstyle[11pt,newpasp,twoside]{article}
\begin{document}
\raggedbottom \lefthyphenmin=3 \righthyphenmin=4 \hyphenpenalty=200

\title{The Stellar Initial Mass Function and Beyond}

\author{Richard B. Larson}

\affil{Yale Astronomy Department, Box 208101, New Haven, CT 06520-8101,
       USA}

\begin{abstract}

   A brief review is given of the basic observed features of the
stellar IMF, and also of some of the theoretical ideas and simulations
that may help to explain these features.  The characteristic stellar
mass of order one solar mass may derive from the typical masses of the
observed star-forming clumps in molecular clouds, and the typical clump
mass may in turn be determined mainly by the Jeans mass, as is true
in many numerical simulations of cloud collapse and fragmentation.
The Salpeter power law that approximates the IMF at higher masses is
less well understood, but it may result from the effects of continuing
gas accretion by the more massive forming stars together with the
effects of interactions and mergers in dense forming stellar systems.
The mass of the most massive star that forms in a cluster is observed
to increase systematically with the mass of the cluster, and the most
massive clusters may form stars that are massive enough to collapse
completely to black holes.

\end{abstract}

\section{Introduction}

   It is almost 50 years since Salpeter (1955) published the first
and still most famous paper on the stellar Initial Mass Function, or
distribution of masses with which stars are formed, and showed that
for masses between 0.4 and 10 solar masses it can be approximated by
a declining power law.  Since then, this `Salpeter law,' or variants
or extensions of it, have been used in countless efforts to model the
properties and evolution of stellar systems.  Recently, for example,
it has been used to model the evolution of galaxies at high redshifts.
Because of its importance in many areas of astronomy, the stellar IMF
has been a subject of intense study for almost five decades, and much
effort has been invested in trying to determine how valid and how
universal the Salpeter law or other approximations to the IMF might be,
and also whether the IMF is variable, depending perhaps on parameters
such as metallicity or location.

   The most important result of all this work can be stated simply:
Salpeter was basically right, and the power-law approximation that
he proposed to the IMF between 0.4 and 10 solar masses has yet to be
conclusively falsified or superseded.  However, we have also learned
that this power law does not apply at all masses: the IMF may not
continue to follow the Salpeter law at the very largest masses, and it
definitely falls below this law at the smallest masses, becoming nearly
flat (in number of stars per unit log mass) below about 0.5 solar
masses.  This flattening of the IMF at low masses implies that most
of the mass in the IMF is contained in stars whose masses are within
an order of magnitude of one solar mass.  Variability of the IMF has
frequently been suggested, and it probably does occur at some level,
but it has yet to be conclusively established.  Extensive reviews of
the literature on the stellar IMF have been presented by Miller \& Scalo
(1979), Scalo (1986, 1998), and Kroupa (2001, 2002); a briefer review
has been given by Larson (1999), and a review with emphasis on young
clusters has been given by Meyer et al.\ (2000).

\section{Summary of The Evidence}

   Evidence concerning the IMF of the field stars in the solar vicinity
has been collected and reviewed by Scalo (1986, 1998) and Kroupa (2001,
2002).  Several proposed approximations to the IMF are all consistent
with the data for masses between about 0.5 and 10 solar masses,
including the original Salpeter law and the lognormal approximation of
Miller \& Scalo (1979).  However, for masses above 10\,M$_\odot$, the
lognormal approximation now seems excluded because it falls off too
steeply; although some results suggest a high-mass slope that is
steeper than the Salpeter law (Scalo 1986, 1998), other results remain
consistent with it (Kroupa 2001, 2002), and the Salpeter law therefore
continues to be widely used as an approximation to the IMF at all masses
above a solar mass.  Below about 0.5 solar masses, however, the field
IMF becomes almost flat in logarithmic units, and it may even decline
in the brown dwarf regime.  If we define the slope $x$ of the IMF by
$dN/d\log m \propto m^{-x}$, the Salpeter slope is $x = 1.35$ (where
the second decimal place has never been significant and the uncertainty
is at least $\pm 0.3$), while the slope at masses below about
0.5\,M$_\odot$ is approximately $x \sim 0$.  Kroupa (2002) suggests an
approximation to the field IMF consisting of three power laws: $x = 1.3$
(essentially the Salpeter law) for $m > 0.5$\,M$_\odot$, $x = 0.3$ (a
nearly flat IMF) for $0.08 < m < 0.5$\,M$_\odot$, and $x = -0.7$ (an IMF
declining toward small masses) for $m < 0.08$\,M$_\odot$.  Many often
inconsistent definitions and notations have been used for the IMF slope,
but some alternative notations that have been widely used are the
parameter $\Gamma = -x$ defined by Scalo and the parameter
$\alpha = 1 + x$ used by Kroupa.

   Scalo (1986) noted a possible tendency for the IMFs of star clusters
to be flatter than the field IMF and more similar to the Salpeter law
for masses above a solar mass, although the data show a large scatter.
The scatter in the IMF slopes of clusters was illustrated graphically
by Scalo (1998), who concluded that either there are real variations in
the IMF or the measurements have very large uncertainties.  Since the
slopes found at masses above 1\,M$_\odot$ scatter roughly equally above
and below the Salpeter value, most researchers have preferred the
conservative interpretation that the data are consistent, within the
large uncertainties, with a universal Salpeter-like IMF for masses above
1\,M$_\odot$.  However the cluster IMFs, like the field IMF, clearly
flatten at smaller masses, and Kroupa's (2002) plot of IMF slope versus
average stellar mass suggests that the slope varies continuously below
1\,M$_\odot$, and that the IMF eventually even declines in the brown
dwarf regime.  Recent studies of young clusters have generally yielded
results consistent with a universal IMF (e.g., Meyer et al.\ 2000;
Luhman et al.\ 2000; Muench et al.\ 2002), but some studies continue to
suggest variability of the IMF; for example, the Taurus region may have
a deficiency of brown dwarfs (Brice\~no et al.\ 2002), while the massive
young clusters in the Galactic Center region may have a relatively
flat upper IMF (Figer et al.\ 1999; Figer 2001; Stolte et al.\ 2003).
Some studies of field populations have also suggested that some old
populations like the Galactic Thick Disk may have a relatively flat
IMF (Reyl\'e \& Robin 2001; Robin \& Reyl\'e 2003).  However, many past
claims of variability have not withstood the test of time, and therefore
the default assumption of a universal IMF is still widely adopted.

   The observed IMF is of interest also because of what it may tell us
about the physics of star formation.  The implications of the IMF for
star formation processes are best illustrated by considering the amount
of mass that goes into stars per unit logarithmic mass interval.  This
function is flatter by one power of mass than the number of stars per
unit log mass discussed above, and it is broadly peaked around one
solar mass, with a slow decline toward larger masses and a much faster
decline below 0.1\,M$_\odot$.  This implies that the processes of star
formation turn most of the star-forming gas into stars whose masses are
within an order of magnitude of one solar mass; although significant
mass also goes into the more massive stars, very little goes into
low-mass stars or brown dwarfs.  Therefore, in terms of where most of
the mass goes, there are two main features of the IMF that need to
be understood: (1) there is a characteristic stellar mass of the order
of one solar mass, and (2) at larger masses the IMF has a power-law tail
similar to the original Salpeter law that also contains significant
mass.

\section{The Origin of the Characteristic Mass}

   What processes might account for a characteristic stellar mass of
the order of one solar mass?  Two basic types of hypotheses have been
proposed to explain this characteristic mass: (1) One possibility is
that stars form from small molecular gas clumps whose masses are similar
to those of the stars that from them; the stars then derive their
typical masses from those of the clumps, whose properties are presumably
determined by cloud fragmentation processes (Larson 1985, 1996, 1999).
(2) Another possibility is that forming stars may continue to accrete
mass indefinitely from an extended medium, and they may `determine their
own masses' by producing outflows that terminate the accretion process
at some stage (Shu, Adams, \& Lizano 1987; Adams \& Fatuzzo 1996).
Neither of these possibilities can yet be definitely excluded, and
both may well play some role in determining stellar masses.  However,
the hypothesis that typical stellar masses are determined by cloud
fragmentation processes is more amenable to direct observational test,
and it appears to receive support from some recent observations.

   Motte, Andr\'e, \& Neri (1998) have made high-resolution millimeter
continuum maps of the rho Ophiuchus star-forming cloud that show many
small, dense, apparently pre-stellar clumps with masses extending down
to well below a solar mass (see also Andr\'e et al.\ 1999; Andr\'e,
Ward-Thompson, \& Barsony 2000).  They found that the mass spectrum of
these clumps closely resembles the stellar IMF, and they suggested on
this basis that the clumps are direct stellar progenitors.  Luhman \&
Rieke (1999) and Motte \& Andr\'e (2001) have noted in addition that the
mass spectrum of these clumps resembles not only the standard stellar
IMF described above, but also the IMF of the young stars in the
$\rho$~Oph cloud itself.  It may be premature to claim that there is
quantitative agreement between the clump mass spectrum and the stellar
IMF, since a similar study of the $\rho$~Oph clumps by Johnstone et al.\
(2000) finds somewhat larger clump masses and questions whether the
smallest clumps found by Motte et al.\ (1998) are gravitationally bound.
Studies of the dense clumps in the Orion clouds by Motte \& Andr\'e
(2001) and Johnstone et al.\ (2001) have also yielded somewhat larger
typical clump masses.  However, the results of all of these studies
are qualitatively very similar, and they all imply the existence of
a typical clump mass of the order of one solar mass.  Thus they all
appear to be consistent with the view that typical stellar masses
derive from the masses of small pre-stellar clumps created by cloud
fragmentation processes.  These results may pose a problem for the
hypothesis that forming stars continue to accrete mass indefinitely from
an extended medium until the accretion is shut off by outflows, since
the observed clump masses are not much larger than those of the stars
that form from them; however, it cannot be ruled out that outflows
still play a significant role in limiting the efficiency with which
stars can form in these clumps.
  
   If the small apparently pre-stellar clumps observed in molecular
clouds are created by cloud fragmentation processes, and if they are
gravitationally bound, then some version of the Jeans criterion, which
balances gravity against thermal pressure, almost certainly plays some
role in determining their typical sizes and masses. The original Jeans
(1929) analysis of the stability of an infinite uniform medium predicted
a minimum mass for a collapsing clump that depends on the temperature
and density of the medium, but the relevance of this analysis has often
been questioned because it was not mathematically self-consistent,
neglecting the overall collapse of the fragmenting medium (Spitzer
1978).  However, rigorous analyses of the stability of equilibrium
configurations that do not undergo overall collapse, such as sheets,
filaments, and disks, yield results that are dimensionally equivalent to
the original Jeans criterion (Larson 1985).  Another type of rigorous
analysis considers the stability of an equilibrium isothermal sphere
with a given temperature and boundary pressure, and derives the size and
mass of a marginally stable or `Bonnor-Ebert' isothermal sphere (Spitzer
1968).  The result is again dimensionally equivalent to the Jeans
criterion, so this type of criterion seems likely to be of quite general
relevance, at least in an approximate dimensional sense.

   If star-forming clumps are produced by turbulent compression in
clouds that have supersonic internal turbulent motions, as suggested by
Larson (1981), then the clumps may initially have boundary pressures
that are comparable to the typical turbulent ram pressure in these
clouds.  Assuming a temperature of 10~K and a typical turbulent cloud
pressure of $3 \times 10^5$ cm$^{-3}$\,K, the mass and radius of a
marginally stable `Bonnor-Ebert' sphere are predicted to be about
0.7\,M$_\odot$ and 0.03\,pc, similar to the typical masses and sizes of
the pre-stellar clumps actually observed in star-forming clouds (Larson
1996, 1998, 1999).  Simulations of supersonic turbulence provide at
least qualitative support for the idea that star-forming clumps are
created by turbulent compression, perhaps typically in clumpy filaments
such as are found quite generally in many types of simulations (e.g.,
Ostriker, Gammie, \& Stone 1999; V\'azquez-Semadeni et al.\ 2000;
Klessen 2001), regardless of the possible role of magnetic fields, but
much more work is needed to clarify the quantitative details of this
process and the initial properties of the collapsing clumps.

   In any case, many simulations of cloud collapse and fragmentation,
both with and without turbulence, have demonstrated the formation of
bound clumps whose masses are typically comparable to the initial Jeans
mass, and in some cases the distribution of clump masses around this
value resembles qualitatively the lower stellar IMF (Larson 1978;
Klessen 2001; Bate, Bonnell, \& Bromm 2002a,b, 2003).  For example,
the bound clumps that appear in Klessen's (2001) simulations of the
collapse and fragmentation of turbulent clouds have a mass spectrum that
is roughly centered on the Jeans mass and that declines at larger masses
but is flatter at lower masses, as observed; similar results have
also been found in the higher-resolution simulations of Bate et al.\
(2002a,b, 2003).  In the latter simulations, significant numbers of
objects are formed with masses all the way down to the opacity limit for
fragmentation, which is about 0.01 solar masses; these very low-mass
objects or `proto-brown dwarfs' are formed by the fragmentation of the
thin dense spiral filaments that appear frequently in these simulations.
Thus ordinary gravitational or Jeans fragmentation, when simulated with
sufficiently high spatial resolution, can yield a considerable range of
masses and may be able to account for the basic observed features of the
lower IMF, including the observed numbers of brown dwarfs.

   Evidence for the existence of a preferred size and mass scale in the
cloud fragmentation process also comes from studies of the clustering
of newly formed stars (Gomez et al.\ 1993; Larson 1995; Simon 1997),
and from studies of the scaling properties of the kinematics (Goodman
et al.\ 1998) and the spatial structure (Williams, Blitz, \& McKee 2000)
of the gas in star-forming molecular clouds.  As discussed by Goodman
et al.\ (1998), all of this evidence is consistent with a picture in
which star-forming clouds are dominated on large scales by turbulence
and chaotic dynamics, possibly self-similar, while on scales smaller
than about 0.1 parsecs the motions become more orderly, and `coherent
cores' can form and collapse to form individual stars or binary systems.
These `coherent cores' are observed to have sizes comparable to the
Jeans length, as might be expected if they form on scales that are small
enough for turbulence to have become unimportant but large enough that
the Jeans criterion is still satisfied.  While many of the details
remain to be clarified, such a picture appears to offer at least a
framework for understanding many features of the formation of low-mass
stars, including the origin of the characteristic stellar mass.

\section{The Origin of the Power-Law Upper IMF}

   The most important effects of star formation on the environment
are produced by the most massive stars, so it is important also to
understand the origin of the power-law upper IMF, which may have a
universal Salpeter-like slope.  The fact that a significant amount of
mass goes into this power law means that the processes that form massive
stars must be fairly efficient, and the possible universality of the
Salpeter slope also suggests that these processes may be universal in
nature and not dependent on local conditions.  Numerous efforts have
been made over the years to explain the Salpeter law, and some of the
types of models that might naturally produce such a power law IMF
have been reviewed by Larson (1991, 1999).  These include (1) clump
coagulation models, which tend under a variety of assumptions to produce
a power-law clump mass spectrum; (2) models involving the continuing
accretional growth of forming stars, which in simple cases can build up
a power-law tail on the IMF; and (3) models postulating self-similar
clustering or cloud structure, which may also naturally produce a
power-law mass spectrum.

   The older literature on the subject contains many models of clump
coagulation that can explain the relatively flat mass spectrum
($x \sim 0.7$) of the CO clumps in molecular clouds, but they do not
directly explain the steeper Salpeter slope ($x \sim 1.35$) without
additional assumptions.  These models do not now seem very relevant
to the stellar IMF, because the CO clumps whose mass spectra they
explain are of relatively low density and are not closely associated
with star formation, unlike the smaller and denser continuum clumps
discussed above.  On the other hand, models postulating the continuing
accretional growth of the most massive forming stars, such as the model
suggested by Zinnecker (1982), now seem more promising when implemented
in the context of realistic simulations of cluster formation.  Zinnecker
(1982) noted that if accreting protostars moving at a typical speed $V$
through a medium of density $\rho$ gain mass at the Bondi-Hoyle rate
$dM/dt \sim G\rho M^2/V^3$, then an initially peaked mass function
gets stretched out toward larger masses and develops a power-law tail
with a slope $x = 1$ that is not greatly different from the Salpeter
slope $x = 1.35$.  A steeper slope would result if the accretion rate
were more strongly dependent on stellar mass, and this might in fact be
expected since the more massive stars tend to form preferentially in the
central regions of clusters where the ambient density $\rho$ is higher
(Larson 1982; Zinnecker, McCaughrean, \& Wilking 1993; Hillenbrand \&
Hartmann 1998; Bonnell \& Davies 1998).  The simulations by Bonnell et
al.\ (1997, 2001; see also Bonnell 2000) of the accretional growth of
stars in a forming cluster confirm this expectation and predict the
development of a power-law tail on the IMF that has a slope similar to
the Salpeter slope.  Thus, continuing accretional growth of the more
massive stars in a forming cluster seems likely to play an important
role in the development of the upper IMF.

   The tendency of stars to form in a hierarchy of groupings, and the
tendency of the more massive stars to form in the larger groupings
(Larson 1982; Zinnecker et al.\ 1993; Elmegreen et al.\ 2000), must
also play some role in the origin of the upper IMF.  A power-law IMF
could result if, for example, stars form in a self-similar clustering
hierarchy, and if the mass of the most massive star that forms in each
subgroup of the hierarchy increases as some power $n$ of the mass of the
subgroup; then a power-law IMF results whose slope is $x = 1/n$ (Larson
1991).  An example of this type of model would be the formation of
stars in a cloud with a fractal mass distribution, such as a cloud
consisting of a fractal network of filaments; for example, if stars
form by the accumulation of matter at the nodes of such a network, and
if they acquire masses proportional to the lengths of the filaments
intersecting there, then the resulting IMF is a power law whose slope
$x$ is equal to the fractal dimension $D$ of the network (Larson 1992).
If this dimension were similar to the fractal dimension $D \sim 1.4$
characterizing the clustering of the T~Tauri stars in the Taurus-Auriga
clouds (Larson 1995), then an IMF with a slope $x \sim 1.4$ would be
predicted by this model.  Other types of fractal models to explain the
Salpeter law, based on different assumptions, have been developed by
Elmegreen (1997, 1999, 2000a).

   A model based only on geometrical concepts is, however, likely to be
too simple since it neglects the effects of the dynamical interactions
that occur in systems of forming stars.  For example, if star formation
involves the accumulation of matter at the nodes of a filamentary
network, then stars as well as gas will accumulate at these nodes and
will interact strongly there.  Interactions among the forming stars can
play an important role in the star formation process itself by helping
to redistribute angular momentum and enabling continuing accretion to
occur (Larson 1990, 2002).  Interactions may play an even more important
role in the formation of massive stars than they do for low-mass stars
because more interactions are likely to be involved (Larson 2002).  An
example of a simple model whereby a sequence of accretion events in a
hierarchy of interacting protostars can lead to a Salpeter-like IMF
was suggested by Larson (1999).  However, the ease with which it is
possible to devise explanations of the Salpeter law shows that this
power law does not by itself place strong constraints on the physics of
massive star formation; numerous theories have been shown to be equally
capable of explaining the Salpeter law.

   Important progress has recently come from high-resolution numerical
simulations of cloud collapse and star formation that include the
effects of continuing gas accretion as well as the effects of
interactions and mergers in dense forming systems of stars (Bonnell
2002; Bonnell \& Bate 2002).  In the detailed simulation of cluster
formation by Bonnell \& Bate (2002), all of these processes occur and
play important roles in the development of the upper stellar IMF.  A
striking result is the occurrence of a runaway increase in the density
of a central core region containing some of the most massive forming
stars, similar to what had been suggested by Larson (1990); this is
accompanied by a runaway growth in the masses of these stars that is
due to continuing gas accretion and mergers, as had been suggested by
Bonnell, Bate, \& Zinnecker (1998).  The collapsing cluster-forming
cloud also develops a  filamentary structure, and groups of stars form
at the nodes of the resulting network of filaments, similar to what
happens in simulations of galaxy formation in cosmological models; these
results may thus reflect quite general features of the evolution of
systems of gas and stars under the influence of gravity and dissipation.

   Like the earlier simulations of cluster formation by Bonnell et al.\
(1997, 2001), the more detailed simulation by Bonnell \& Bate (2002)
yields a power-law upper IMF that is very similar to the Salpeter law.
Because many processes occur in this simulation and the dynamics
becomes highly complex and chaotic, the reason for this result is not
immediately apparent, but almost certainly a combination of processes
is involved, including all of those discussed above and perhaps others
as well.  It is nevertheless worth noting, as these authors point out,
that the basic physics included in this simulation is very simple --
only isothermal gas dynamics, Newtonian gravity, and the merging of
protostars at close distances, but magnetic fields are not included,
and neither are any effects of stellar evolution or radiative feedback.
Because isothermal gas dynamics and gravity introduce no new mass
scale larger than the Jeans mass, the dynamics of the system becomes
nearly scale-free on large scales, so it is not surprising that a
power-law upper IMF should result.  However, it is not obvious why the
slope should be very similar to the Salpeter law, and more work will
again be needed to clarify the processes involved.

\section{The Most Massive Objects}

   Clearly, it is particularly important to understand better how the
most massive stars are formed.  It is also relevant at a meeting on
``Star Formation Across the Stellar Mass Spectrum'' to ask how far the
stellar mass spectrum extends, and whether there is an upper limit on
stellar masses.  If a Salpeter IMF were to extend indefinitely to very
large masses, this would imply that there is no mass limit and that
stars of larger and larger mass form in systems of larger and larger
size.  However, the Salpeter law cannot continue indefinitely, because
this would imply the existence of some stars in our Galaxy with masses
of 1000 solar masses or more, which are not observed (Elmegreen 2000b).
The formation of massive stars remains, unfortunately, a very poorly
understood subject because massive stars form in very complex
environments that are difficult to study observationally and model
theoretically (Evans 1999; Garay \& Lizano 1999; Stahler, Palla, \& Ho
2000).  Thus it may be of interest to see what can be learned just from
the phenomenology of massive star formation, and from observed trends
such as the dependence of maximum stellar mass on system mass.

   The most massive stars form near the centers of the densest and
most massive clusters (Zinnecker et al.\ 1993; Hillenbrand \& Hartmann
1998; Elmegreen et al.\ 2000; Clarke, Bonnell, \& Hillenbrand 2000),
and the mass of the most massive star present also tends to increase
systematically with the mass of the cluster (Larson 1982; Testi, Palla,
\& Natta 1999; Testi 2003).  This is illustrated by comparing the
$\rho$~Oph cluster, which has a mass of about 100\,M$_\odot$ and whose
most massive star has a mass of about 8\,M$_\odot$ (Wilking, Lada, \&
Young 1989), with the Orion Nebula cluster, which has a mass of about
2000\,M$_\odot$ and whose most massive star, located in the Trapezium at
its center, has a mass of about 40\,M$_\odot$ (Hillenbrand \& Hartmann
1998).  Even the Orion Nebula cluster is only a modest example of
massive star formation, however; for example, the Arches and Quintuplet
clusters near the Galactic Center have masses of at least $10^4$ solar
masses and contain stars with masses well over 100\,M$_\odot$ (Figer et
al.\ 1999).  The most luminous and massive star known may be the `Pistol
star' in the Quintuplet cluster; the mass of this star is estimated to
be at least 150\,M$_\odot$, and the mass of the associated cluster is
about $2 \times 10^4$\,M$_\odot$ (Figer et al.\ 1998, 1999; Figer \&
Kim 2002).  A similarly luminous star, LBV\,1860-20, has been found
in a similarly massive cluster that is not near the Galactic Center
(Eikenberry et al.\ 2001).  The R136 cluster in the Large Magellanic
Cloud, which has a mass of about $5 \times 10^4$\,M$_\odot$, also
contains stars with masses up to about 150\,M$_\odot$ (Hunter et al.\
1995; Massey \& Hunter 1998).

   The examples mentioned above, namely the $\rho$~Oph cluster, the
Orion Nebula cluster, and the Quintuplet and R136 clusters, together
with a number of other young systems for which data were collected by
Larson (1982), define a rough power-law correlation between the mass of
the most massive star and the mass of the associated cluster which is
approximately $M_{\rm star,\,max} \sim 1.2\,M_{\rm cluster}^{0.45}$.
From these very limited data it is not clear that any limiting stellar
mass has yet been reached, since the mass of the most massive star
increases more or less continually with cluster mass up to the most
massive young clusters known in our Galaxy and the LMC.  However, the
dependence of maximum stellar mass on cluster mass found here is not
consistent with a simple extension of the Salpeter law to very large
masses, since this would predict that the maximum stellar mass
should vary as $M_{\rm cluster}^{0.74}$, rather than only as
$M_{\rm cluster}^{0.45}$.  If the data used here are reliable and
representative, they suggest that the upper IMF eventually becomes
steeper than the Salpeter law and approaches a slope that is closer
to $x = 1/0.45 = 2.2$ at very large masses.  Thus the IMF may not be
truly self-similar for the most massive stars but may have a slope
that increases with mass, possibly reflecting an increasing difficulty
in forming progressively more massive stars that is due to the effects
of radiation pressure and winds. 

   Do stars with even larger masses form in more massive clusters, such
as young globular clusters?  It has long been suspected that the massive
centrally condensed globular cluster M15 has a central black hole with a
mass of the order of 1000\,M$_\odot$, and this possibility is supported
by recent radial velocity measurements of stars only a few tenths of an
arcsecond from the center (Gebhardt et al.\ 2000; Gerssen et al.\ 2001).
According to Gebhardt et al.\ (2000), this evidence is consistent with
a black hole with a mass of $\sim 2000$\,M$_\odot$, although other
interpretations are not excluded, and the STIS measurements of Gerssen
et al.\ (2001) provide further support for an increase in stellar
velocity dispersion toward the center.  If M15 does indeed contain such
a central black hole, one explanation could be that a similarly massive
star formed at the center of this cluster and then collapsed to a black
hole.  Stars with masses larger than 250\,M$_\odot$ are predicted to
collapse completely to black holes at the end of their lifetimes,
provided that they do not first lose most of their mass in winds (Fryer,
Woosley, \& Heger 2001), and it may be relevant in this case that mass
loss is expected to be relatively unimportant for metal-poor stars
(Baraffe, Heger, \& Woosley 2001) and that M15, with ${\rm [Fe/H]} =
-2.0$, is one of the most metal-poor clusters known.  M15 might thus
have provided a particularly favorable environment for the formation of
a very massive star that could have collapsed to a massive black hole.

   Other centrally condensed systems are also known that contain central
black holes: supermassive black holes are found at the centers of most,
if not all, galaxy bulges, and the masses of these nuclear black holes
increase systematically with the mass of the bulge (Kormendy \&
Richstone 1995; Kormendy \& Gebhardt 2001).  If M15 contains a black
hole of mass $\sim 2000$\,M$_\odot$, it falls on the same relation
between black hole mass and system mass that was found by Kormendy \&
Richstone (1995) for galaxy bulges, since these authors found that the
mass of the central black hole is typically about 0.002 times the mass
of the system, which is also approximately true for M15.  The updated
reviews of Kormendy \& Gebhardt (2001) and McLure \& Dunlop (2002) find,
for a much larger sample of galaxies, a very similar relation between
black hole mass and bulge mass, with a slightly smaller value of 0.0013
for the typical ratio of black hole mass to bulge mass.

   Could there be a connection between the formation of massive stars at
the centers of clusters and the formation of supermassive black holes at
the centers of galaxies?  In both cases, the mass of the central object
increases systematically with the mass of the system, and this suggests
that similar processes may be involved in building up the central object
(Larson 2002).  For example, runaway accretion and coalescence processes
like those discussed above could play a role not only in forming massive
stars in clusters but in building supermassive black holes in galactic
nuclei.  The relation between central object mass and system mass
is not identical in the two cases, however, since the mass of the
most massive star in a cluster increases with system mass only as
$1.2\,M_{\rm system}^{0.45}$, while the mass of the central black
hole in a galaxy bulge increases more nearly linearly with system mass
and is approximately equal to $0.0015\,M_{\rm system}$.  These two
relations intersect at a system mass of about $2 \times 10^5$\,M$_\odot$
and a central object mass of about 300\,M$_\odot$, suggesting that a
transition may occur at about this point from a regime of normal massive
star formation in systems of smaller mass to a regime of black hole
formation and growth in systems of larger mass.  It is consistent with
this possibility that the transition appears to occur at a maximum
object mass of around 300\,M$_\odot$, since this is approximately the
same as the mass of 250\,M$_\odot$ above which stars are predicted to
collapse completely to black holes; this suggests that systems more
massive than a few times $10^5$\,M$_\odot$ may sometimes make stars
massive enough to collapse completely to black holes.

   Once a massive black hole has formed, the possibilities for the
further growth in mass of this object are increased because such a
black hole, unlike a massive star, is a permanent feature of the system
that can only grow in mass by any continuing interaction with its
surroundings.  For example, a central black hole might continue to
accrete matter from neighboring massive stars in a central subcluster
whose members might otherwise have merged.  Such processes might have
contributed to the mass of the possible black hole in M15, since the
maximum stellar mass predicted by the above relation for a cluster with
the mass of M15 is only about 600\,M$_\odot$, and this suggests that
some further growth in mass might have been required.  In the case of
galaxy bulges, which are built up at least partly by mergers of smaller
systems, the mergers may lead to the continuing growth of the nuclear
black holes through the accretion of gas, stars, or other black holes
that become concentrated at the center during each merger, and such
processes might account for the existence of a relation between black
hole mass and bulge mass (Cattaneo, Haehnelt, \& Rees 1999;
Kauffmann \& Haehnelt 2000; Larson 2000).

   Even without gas dynamics, runaway stellar mergers can occur in
massive and dense young star clusters when they experience a stellar
dynamical `core collapse', and this can lead to the building up of very
massive stars and black holes at their centers.  Portegies Zwart et al.\
(1999, 2002) and Figer \& Kim (2002) suggest that stellar mergers might
account for the existence of such very massive stars as the Pistol star,
and Ebisuzaki et al.\ (2001) and Portegies Zwart \& McMillan (2002)
suggest that black holes with masses of hundreds to thousands of solar
masses may be built up by similar processes at the centers of massive
young clusters.  X-ray observations provide evidence that such
`intermediate mass black holes' exist in some very luminous young
clusters in starburst galaxies (Ebisuzaki et al.\ 2001; Portegies Zwart
\& McMillan 2002), and this evidence supports the possibility that
massive black holes can indeed form at the centers of massive young
clusters.  Portegies Zwart \& McMillan (2002) speculate further that the
accumulation of such clusters at the centers of galaxies, accompanied
by the merging of their black holes, could lead to the building up of
supermassive black holes in galactic nuclei, and they argue that central
objects with masses of the order of 0.1 percent of the system mass
should result quite generally from such processes.

\section{Summary}

   Much evidence indicates that there are two basic general features
of the stellar IMF that need to be understood: there is a characteristic
stellar mass of the order of one solar mass, and the IMF also has a
power-law tail toward larger masses that is similar to the original
Salpeter law.  Variability of the IMF has often been suggested, but
has not been conclusively established.  The characteristic mass may
derive from the typical masses of the small, dense, apparently
pre-stellar clumps that have been observed in a number of star-forming
clouds.  In many simulations of cloud collapse and fragmentation, the
typical masses of the clumps formed are similar to the Jeans mass,
regardless of whether turbulence plays an important role; magnetic
fields also do not appear to make a major difference.  Gravitational
fragmentation produces in addition significant numbers of proto-brown
dwarfs by the fragmentation of thin dense spiral filaments, and it may
thus be able to account for the basic observed features of the lower
stellar IMF.  The Salpeter power-law that approximates the IMF at
higher masses is less well understood, but it probably results from
a combination of processes including the effects of continuing gas
accretion by the more massive forming stars and the effects of
interactions and mergers in dense forming systems of stars.  Simulations
of the formation of star clusters show that the dynamics of these
accretional growth processes becomes almost scale-free at large masses,
and that these processes can produce a power-law upper IMF that is very
similar to the Salpeter law.

   Observations also indicate that the mass of the most massive star
that forms in a cluster increases systematically with the mass of the
cluster, although not as rapidly as is predicted by the Salpeter law,
and there is no clear indication of an upper stellar mass limit.  Some
of the most massive known clusters, including the globular cluster M15
and some luminous young clusters in starburst galaxies, may contain
massive black holes that originated from the collapse of very massive
stars that formed near the centers of these systems.  The observed
dependence of maximum object mass on system mass suggests that stellar
systems more massive than a few times $10^5$ solar masses may sometimes
make stars massive enough to collapse completely to black holes.

\end{document}